\documentclass[useAMS,usenatbib,usegraphicx]{mn2e}
\usepackage{epsfig}
\usepackage{appendix}
\usepackage{booktabs}
\usepackage{upgreek}
\usepackage{url}
\usepackage[usenames]{color}
\newcommand{\ion}[2]{{#1}\,{\sc #2}}

\usepackage{graphicx}
\usepackage{rotating}
\usepackage{color}
\usepackage{pifont}
\usepackage{longtable}
\usepackage{pdflscape,lscape}
\usepackage{natbib}

\title[
The magnetic field geometry of CPD\,$-57^{\circ}$3509
]{
Magnetic field geometry and chemical abundance distribution of the He-strong star CPD\,$-$57$^{\circ}$3509
}

\author[
Hubrig et al.
]{
S.~Hubrig$^1$\thanks{E-mail: shubrig@aip.de},
N.~Przybilla$^2$,
H.~Korhonen$^3$,
I.~Ilyin$^1$,
M.~Sch\"oller$^4$,
S.~P.~J\"arvinen$^1$,
\and M.-F.~Nieva$^2$,
R.-D.~Scholz$^1$,
S.~Kimeswenger$^{5,2}$,
M.~Ramolla$^6$,
A.~F.~Kholtygin$^7$,
\and M.~Briquet$^{1,8}$\\
$^1$Leibniz-Institut f\"ur Astrophysik Potsdam (AIP), An der Sternwarte~16, 14482~Potsdam, Germany\\
$^2$Institut f\"ur Astro- und Teilchenphysik, Universit\"at Innsbruck, Technikerstr.~25/8, 6020~Innsbruck, Austria\\
$^3$Dark Cosmology Centre, Niels Bohr Institute, University of Copenhagen, Juliane Maries Vej~30, 2100~Copenhagen~Ø, Denmark\\
$^4$European Southern Observatory, Karl-Schwarzschild-Str.~2, 85748~Garching, Germany\\
$^5$Instituto de Astronom\'{i}a, Universidad Cat\'{o}lica del Norte, Avenida Angamos 0610, Casilla 1280, Antofagasta, Chile\\
$^6$Astronomisches Institut, Ruhr-Universit{\"a}t Bochum, Universit{\"a}tsstr.~150, 44801 Bochum, Germany\\
$^7$Saint-Petersburg State University, Universitetskij pr. 28, Saint-Petersburg 198504, Russia \\
$^8$Institut d'Astrophysique et de G\'eophysique, Universit\'e de Li\`ege, All\'ee~du~6~Ao\^ut, B\^at.~B5c, 4000~Li\`ege, Belgium
}

\begin{document}

\def\teff{${T}_{\rm eff}$}
\def\kms{{km\,s}$^{-1}$}
\def\logg{$\log g$}
\def\ksi{$\xi_{\rm t}$}
\def\micro{$\xi_{\rm t}$}
\def\macro{$\zeta_{\rm RT}$}
\def\rad{$v_{\rm r}$}
\def\vsini{$v\sin i$}
\def\ebv{$E(B-V)$}

\date{Accepted ... Received ...; in original form ...}

\pagerange{\pageref{firstpage}--\pageref{lastpage}} \pubyear{2002}

\maketitle

\label{firstpage}

\begin{abstract}
The magnetic field of CPD\,$-57^{\circ}$3509 was recently detected
in the framework of the BOB (B fields in OB stars) collaboration.
We acquired low-resolution spectropolarimetric observations of 
CPD\,$-57^{\circ}$3509 with 
FORS\,2 and high-resolution UVES observations randomly distributed over a few months
to search for periodicity, to study the magnetic field geometry,
and to determine the surface distribution of silicon and helium. We also obtained 
supplementary photometric observations at a timeline similar 
to the spectroscopic and spectropolarimetric observations.
A period of 6.36\,d was detected in the measurements of the mean longitudinal
magnetic field. 
A sinusoidal fit to our measurements
allowed us to constrain the magnetic field geometry and estimate the dipole strength in the 
range of 3.9--4.5\,kG.
Our application of the Doppler imaging technique revealed the presence of
\ion{He}{i} spots located around the magnetic poles, with a strong concentration at the positive 
pole and a weaker one around the negative pole. 
In contrast, high concentration \ion{Si}{iii} spots
are located close to the magnetic equator. 
Further, our analysis of the spectral variability of CPD\,$-57^{\circ}$3509
on short time scales indicates
distinct changes in shape and position of line profiles possibly caused by the presence of $\beta$~Cep-like
pulsations.  A small periodic variability in
line with the changes of the magnetic field strength is clearly seen in the photometric data.
\end{abstract}


\section{Introduction}  
\label{sect:intro}

   Recently, \citet{Przybilla2016} presented a firm detection of a mean longitudinal magnetic field
   of kG order in the early-B type star CPD\,$-57^{\circ}$3509, previously studied in a spectroscopic survey of 
massive stars in NGC\,3293 by \citet{Evans2005}.
In their work, the authors focussed on the investigation of abundances, a model atmosphere, and the evolutionary
state of the target in detail.
The quantitative spectroscopic analysis of this star with the observed rather low $v\,\sin\,i$-value of 35\,km\,s$^{-1}$
 yielded an effective temperature and a 
   logarithmic surface gravity of $23\,750\pm250$\,K and $4.05\pm0.10$,
   respectively, and a surface helium fraction of $0.28\pm0.02$ by
   number (see also \citealt{Maeder2014}). The surface abundances of C, N, O, Ne, S, and Ar were found to be
   compatible with the cosmic abundance standard \citep{Nieva2012}, whereas Mg, Al, Si, and Fe were
   depleted by about a factor of 2. It was suggested that such an abundance pattern can be
   understood as the consequence of a fractionated stellar wind.
   Importantly, CPD\,$-57^{\circ}$3509 with an elapsed main-sequence life time of about 50\% 
 has evolved significantly away from the zero-age main sequence and appears to 
   be one of the most evolved He-strong stars known with an independent age constraint 
   due to its cluster membership. 
Since the evolution of the magnetic field geometry across the main sequence in massive B-type stars is 
not well sampled in comparison to studies
of magnetic fields of late-type Bp and intermediate-mass Ap stars, a detailed study of the magnetic field 
configuration and the surface chemical inhomogeneities in a significantly evolved magnetic He-strong star is of
particular interest. Knowledge of the evolution of the magnetic field geometry, especially of the distribution 
of the obliquity angle
$\beta$ (the orientation of the magnetic axis with respect
to the rotation axis) is essential to understand the physical
processes taking place in these stars and the origin of their
magnetic fields. Further, although it is generally assumed that CPD$-$57$^{\circ}$3509 is a member of
the open cluster NGC\,3293, no careful study of its membership involving new data from Gaia was carried out yet.
We review the membership status involving the available Gaia data in Appendix~\ref{sect:appA}.

\begin{table*}
\caption{
Logbook of the FORS\,2 polarimetric observations of CPD\,$-$57$^{\circ}$3509, including 
the modified Julian date of mid-exposure followed by the
achieved signal-to-noise ratio in the Stokes~$I$ spectra around 5000\,\AA{},
and the measurements of the mean longitudinal magnetic field using the 
Monte Carlo bootstrapping test, only for the hydrogen lines and for all lines.
In the last columns, we present the results of our measurements using the null spectra for the set
of hydrogen and all lines and the phases calculated
relative to a zero phase corresponding to a positive field extremum at MJD56983.5063 for the set with
hydrogen lines assuming the rotation period $P_{\mathrm{rot}}=6.36093$\,d and at MJD56983.5396 for 
the set with all lines assuming the rotation period $P_{\mathrm{rot}}=6.36255$\,d.
All quoted errors are 1$\sigma$ uncertainties. 
Please note that the measurement marked by an asterisk on MJD\,57027.3200 suffered from the fact that
the guide star was lost several times due to thick clouds. 
}
\label{tab:log_meas}
\centering
\begin{tabular}{lrr@{$\pm$}rr@{$\pm$}rr@{$\pm$}rr@{$\pm$}rcc}
\hline
\hline
\multicolumn{1}{c}{MJD} &
\multicolumn{1}{c}{S/N} &
\multicolumn{2}{c}{$\left< B_{\rm z}\right>_{\rm hyd}$} &
\multicolumn{2}{c}{$\left< B_{\rm z}\right>_{\rm all}$} &
\multicolumn{2}{c}{$\left< N_{\rm z}\right>_{\rm hyd}$} &
\multicolumn{2}{c}{$\left< N_{\rm z}\right>_{\rm all}$} &
\multicolumn{1}{c}{$\varphi_{\rm hyd}$}&
\multicolumn{1}{c}{$\varphi_{\rm all}$}\\ 
&
 &
 \multicolumn{2}{c}{[G]} &
 \multicolumn{2}{c}{[G]}  &
 \multicolumn{2}{c}{[G]} &
 \multicolumn{2}{c}{[G]} &
& 
\\
\hline
     56695.2463&1381 & $-$287&  126  & $-$23 & 60&$-$377& 139  &$-$101&  64 &0.683& 0.689\\
     56696.2876&1826 &    694&  108  &  539  & 51&$-$116& 104  &     1&  48 &0.846& 0.853\\ 
     56810.0089&2025 &  $-$19&   71  &   88  & 54&$-$28 &  86  &$-$45 &  59 &0.725& 0.726\\
     56810.9983&2348 &    979&   68  &  920  & 48&$-$108&  77  &    2 &  50 &0.880& 0.882\\
     56978.3148&1397 &    966&  159  &  835  & 88&$-$70 & 166  &   31 &  94 &0.184& 0.179\\ 
     57012.2145& 943 & $-$423&  293  &$-$650 &202&$-$162& 235  &   23 & 220 &0.513& 0.507\\ 
     57024.3376&1125 & $-$588&  237  &$-$815 &137&  230 & 192  &  211 & 121 &0.419& 0.412\\ 
     57025.2844&1646 & $-$590&  127  &$-$557 & 75&$-$87 & 132  &    5 &  74 &0.568& 0.561\\ 
     57026.2200&1127 &     24&  174  &    32 &105&  178 & 163  &  107 & 100 &0.715& 0.708\\ 
     57027.3200$^{\ast}$& 665 & $-$106&  384  & $-$62 &199&  237 & 331  &  145 & 202 &0.888& 0.881\\
     57030.2079&1406 & $-$627&  138  &$-$444 & 94&  249 & 157  &   89 & 105 &0.342& 0.335\\
     57031.3116&1507 &$-$1052&  121  &$-$847 & 73&$-$65 & 141  &$-$82 &  81 &0.515& 0.508\\
     57039.3509&1472 &    347&  140  &   272 & 77&$-$6  & 134  &   35 &  71 &0.779& 0.772\\
     57057.2937&1835 & $-$666&  107  &$-$508 & 64&$-$87 & 114  &$-$18 &  63 &0.600& 0.592\\
     57059.1600&1217 &    709&  173  &   791 & 98&$-$169& 188  &$-$14 & 104 &0.893& 0.885\\
     57060.1721&1318 &   1209&  157  &  1122 &104&$-$8  & 170  &$-$20 &  99 &0.053& 0.044\\
     57069.0658&1343 &$-$1021&  164  & $-$787&110&  4   & 175  &  119 & 118 &0.451& 0.442\\
     57070.0817&1175 & $-$818&  158  & $-$606&136&  257 & 195  & $-$61& 132 &0.610& 0.602\\
     57071.0569&1086 &    127&  202  &    212&118&   30 & 182  & $-$74& 105 &0.764& 0.755\\
     57099.0931&1826 &    582&   99  &    671& 62& $-$75& 101  & $-$33&  61 &0.171& 0.162\\
\hline
\end{tabular}
\end{table*}

In order to characterise the properties of CPD\,$-57^{\circ}$3509 in detail, we obtained time-series spectroscopy and
spectropolarimetry with the FOcal Reducer low dispersion 
Spectrograph (FORS\,2; \citealt{Appenzeller1998}) and the UV-Visual Echelle Spectrograph (UVES; \citealt{UVES}) to constrain 
the magnetic field geometry and reconstruct the distribution
of silicon and helium on the stellar surface using Doppler Imaging (DI). Further, we carried out photometric 
observations using the 40\,cm Bochum
Monitoring Telescope (BMT; \citealt{Ramolla2013}) of the Cerro Armazones Observatory.
In the first part of the paper we report on the 
period determination using magnetic and photometric data. 
In the second part we present the results of the application of the 
Doppler Imaging technique followed by the discussion of
the surface element distribution with respect to the 
magnetic field configuration.

\section{Observations and magnetic field measurements}
\label{sect:obs}

Fifteen FORS\,2 spectropolarimetric observations of CPD\,$-$57$^{\circ}$3509 were obtained
in the framework of the ESO programme 094.D-0355 from 2014 November 17 to 2015 February 18, and further five 
within the framework of the ESO Large Programme 191.D-0255 during 2014 February and June, and 2015 March.
The FORS\,2 multi-mode instrument is equipped with polarisation analysing optics
comprising super-achromatic half-wave and quarter-wave phase retarder plates,
and a Wollaston prism with a beam divergence of 22$\arcsec$ in standard
resolution mode. 
We used the GRISM 600B and the narrowest available slit width
of 0$\farcs$4 to obtain a spectral resolution of $R\sim2000$.
The observed spectral range from 3250 to 6215\,\AA{} includes all Balmer lines,
apart from H$\alpha$, and numerous helium lines.
For the observations, we used a non-standard readout mode with low 
gain (200kHz,1$\times$1,low), which provides a broader dynamic range, hence 
allowing us to reach a higher signal-to-noise ratio (S/N) in the individual spectra.
The exposure time for each subexposure 
accounted for 5\,min.
Each observation consisted of eight subexposures over approximately one hour including overheads.

Our first description of the assessment of longitudinal magnetic field
measurements using FORS\,1/2 spectropolarimetric observations was presented 
in several previous works (e.g.\ \citealt{Hubrig2004a,Hubrig2004b}, 
and references therein).
To minimize the cross-talk effect
and to cancel errors from 
different transmission properties of the two polarised beams,
a sequence of subexposures at the retarder
position angles
$-$45$^{\circ}$$+$45$^{\circ}$,
$+$45$^{\circ}$$-$45$^{\circ}$,
$-$45$^{\circ}$$+$45$^{\circ}$,
etc.\ is usually executed during the observations. Moreover, the reversal of the quarter wave 
plate compensates for fixed errors in the relative wavelength calibrations of the two
polarised spectra.
According to the FORS User Manual, the $V/I$ spectrum is calculated using:
\begin{equation}
\frac{V}{I} = \frac{1}{2} \left\{ 
\left( \frac{f^{\rm o} - f^{\rm e}}{f^{\rm o} + f^{\rm e}} \right)_{-45^{\circ}} -
\left( \frac{f^{\rm o} - f^{\rm e}}{f^{\rm o} + f^{\rm e}} \right)_{+45^{\circ}} \right\}
\end{equation}
where $+45^{\circ}$ and $-45^{\circ}$ indicate the position angle of the
retarder waveplate and $f^{\rm o}$ and $f^{\rm e}$ are the ordinary and
extraordinary beams, respectively. 
Rectification of the $V/I$ spectra was
performed in the way described by \citet{Hubrig2014a}.
Null profiles, $N$, are calculated as pairwise differences from all available 
$V$ profiles.  From these, 3$\sigma$-outliers are identified and used to clip 
the $V$ profiles.  This removes spurious signals, which mostly come from cosmic
rays, and also reduces the noise. A full description of the updated data 
reduction and analysis will be presented in a separate paper (Sch\"oller et 
al., in preparation; see also \citealt{Hubrig2014a}).
The mean longitudinal magnetic field, $\left< B_{\rm z}\right>$, is 
measured on the rectified and clipped spectra based on the relation 
following the method suggested by \citet{Angel1970}:

\begin{eqnarray} 
\frac{V}{I} = -\frac{g_{\rm eff}\, e \,\lambda^2}{4\pi\,m_{\rm e}\,c^2}\,
\frac{1}{I}\,\frac{{\rm d}I}{{\rm d}\lambda} \left<B_{\rm z}\right>\, ,
\label{eqn:vi}
\end{eqnarray} 

\noindent 
where $V$ is the Stokes parameter that measures the circular polarization, $I$
is the intensity in the unpolarized spectrum, $g_{\rm eff}$ is the effective
Land\'e factor, $e$ is the electron charge, $\lambda$ is the wavelength,
$m_{\rm e}$ is the electron mass, $c$ is the speed of light, 
${{\rm d}I/{\rm d}\lambda}$ is the wavelength derivative of Stokes~$I$, and 
$\left<B_{\rm z}\right>$ is the mean longitudinal (line-of-sight) magnetic field.

The longitudinal magnetic field was measured in two ways: using the entire spectrum
including all available lines, or using exclusively hydrogen lines.
Furthermore, we have carried out Monte Carlo bootstrapping tests. 
These are most often applied with the purpose of deriving robust estimates of standard errors. 
The measurement uncertainties obtained before and after the Monte Carlo bootstrapping tests were found to be 
in close agreement, indicating the absence of reduction flaws. 
The results of our magnetic field measurements, those for the entire spectrum
and only the hydrogen lines are presented in 
Table~\ref{tab:log_meas}, where we also include in the first rows information 
about the previous magnetic field measurements presented by \citet{Przybilla2016}. The last row
shows an additional measurement obtained on 2015 March 18 in the framework of the BOB ESO Large Programme 191.D-0255.
The rotation phases presented in the last columns of Table~\ref{tab:log_meas} were calculated 
using the ephemeris 
determined from our period search described in Sect.~\ref{sect:mag}.

\begin{table}
  \caption[]{Logbook of the UVES observations. The phases were calculated
relative to a zero phase corresponding to a positive field extremum at MJD56983.5396 for the set with
all lines assuming the rotation period $P_{\mathrm{rot}}=6.36255$\,d}. 
  \label{tab:obs}
\centering
  \begin{tabular}{cccc}
    \hline
    \hline
    \noalign{\smallskip}
\multicolumn{1}{c}{MJD} &
\multicolumn{1}{c}{Date} &
\multicolumn{1}{c}{Phase} &
\multicolumn{1}{c}{S/N} \\
    \noalign{\smallskip}
    \hline
    \noalign{\smallskip}
    56997.27269 & 2014/12/06 & 0.159 & 175 \\
    57000.30922 & 2014/12/09 & 0.636 & 223 \\
    57023.20634 & 2015/01/01 & 0.234 & 189 \\
    57025.19286 & 2015/01/03 & 0.547 & 168 \\
    57047.23708 & 2015/01/25 & 0.011 & 249 \\
    57052.20926 & 2015/01/30 & 0.793 & 191 \\
    57055.07998 & 2015/02/02 & 0.244 & 228 \\
    57056.15139 & 2015/02/03 & 0.412 & 255 \\
    57082.06466 & 2015/03/01 & 0.485 & 244 \\
    57087.14368 & 2015/03/06 & 0.284 & 227 \\
    \noalign{\smallskip}
    \hline
  \end{tabular}
\end{table}

High resolution spectra of CPD\,$-57^{\circ}$3509 were obtained at the European Southern Observatory with the UVES 
high resolution spectrograph. The observations with an exposure time of about 40\,min were obtained 
from 2014 December 6 to 2015 March 6 using the 
standard wavelength setting for dichroic mode (DIC-2\,437+760) with the  0.4\arcsec{} slit in the blue arm 
giving a spectral resolving 
power of $R\sim80,000$ and the 0.3\arcsec{} slit in the red arm to achieve a resolution of $R\sim110,000$.
The wavelength coverage was 3730--9390\,\AA{} with a gap between 5000 and 5700\,\AA{}. 
Initially, we asked for 20 observations in service
mode, but, unfortunately, only ten observations were executed. The summary of the UVES spectroscopic 
observations is presented in Table~\ref{tab:obs}. The table gives the Modified Julian date at the middle of each
observation, the observing date, the rotational phase, and the S/N.
The S/N is given per resolution element and is measured from the spectral region around 
4630\,\AA{}.
All observations were phased using the ephemeris ${\rm HJD} = 2\,456\,983.5396 + 6.36255\times E$, 
referring to the time of the maximum positive magnetic field.
Advantageously, the phase distribution
of the obtained spectra appears suitable for a Doppler imaging analysis to study the surface distribution
of silicon and helium, which exhibit the most distinct spectrum variability in He-strong stars.   

Supplementary photometric observations were carried out at the
BMT.
Taken in 2014 April and May and during
2015 March 31 to April 4, the photometry covers a timeline similar
to the spectroscopic and spectropolarimetric observations.
Johnson~$B$ and $V$ filters were used and a 3$\times$3 dithering pattern of
observations were
obtained each night to improve the sampling and photometric accuracy. Seven
photometrically invariable cluster stars (numbers 3, 44, 45, 51, 58, 97,
and 110 of \citealt{Baume2003})
in the vicinity of CPD\,$-57^{\circ}$3509 were used for comparison
in differential photometry. They covered a
magnitude range of 9\fm4 $\le$ (m$_{\mathrm V}$,m$_{\mathrm B}$) $\le$
12\fm9 and colors
from 0\fm06 $\le$ (B-V) $\le$ 0\fm27, thus well enclosing the target
and minimizing filter effects. The rms between the comparison stars were
$\sigma_{\mathrm V} = 0\fm0047$ and $\sigma_{\mathrm B} = 0\fm0062$.
The photometric data are summarized in Tables~\ref{tab:photometry_B} and \ref{tab:photometry_V}, presented in Appendix~\ref{sect:appB}.
Our photometric zero values (median of the 2014 data points) for CPD\,$-57^{\circ}$3509 are
$B=10.80\pm0.01$ and $V=10.69\pm0.01$.

\section{Period determination}
\label{sect:mag}

\begin{figure*}
\centering
\includegraphics[width=0.45\textwidth]{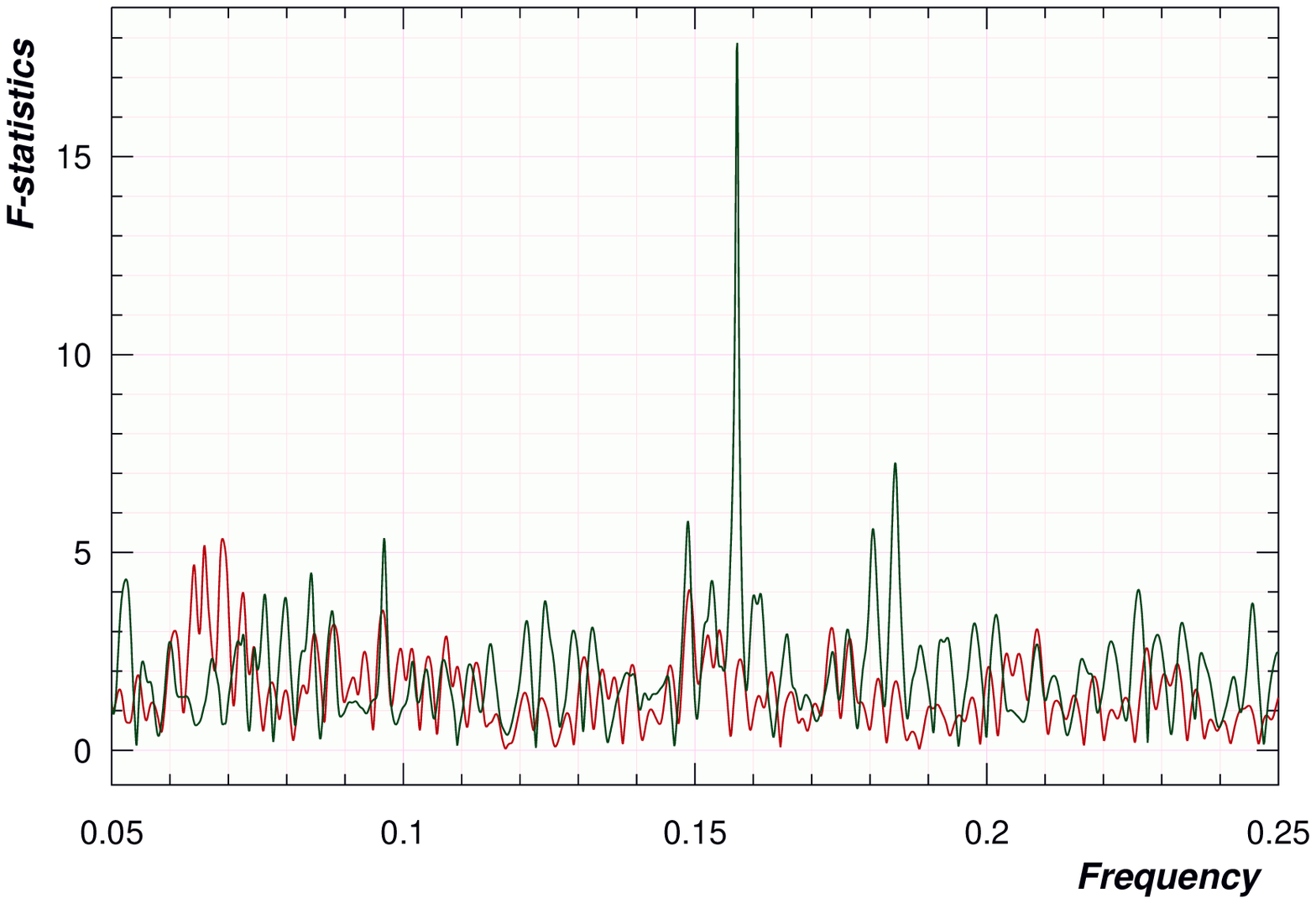}
\includegraphics[width=0.45\textwidth]{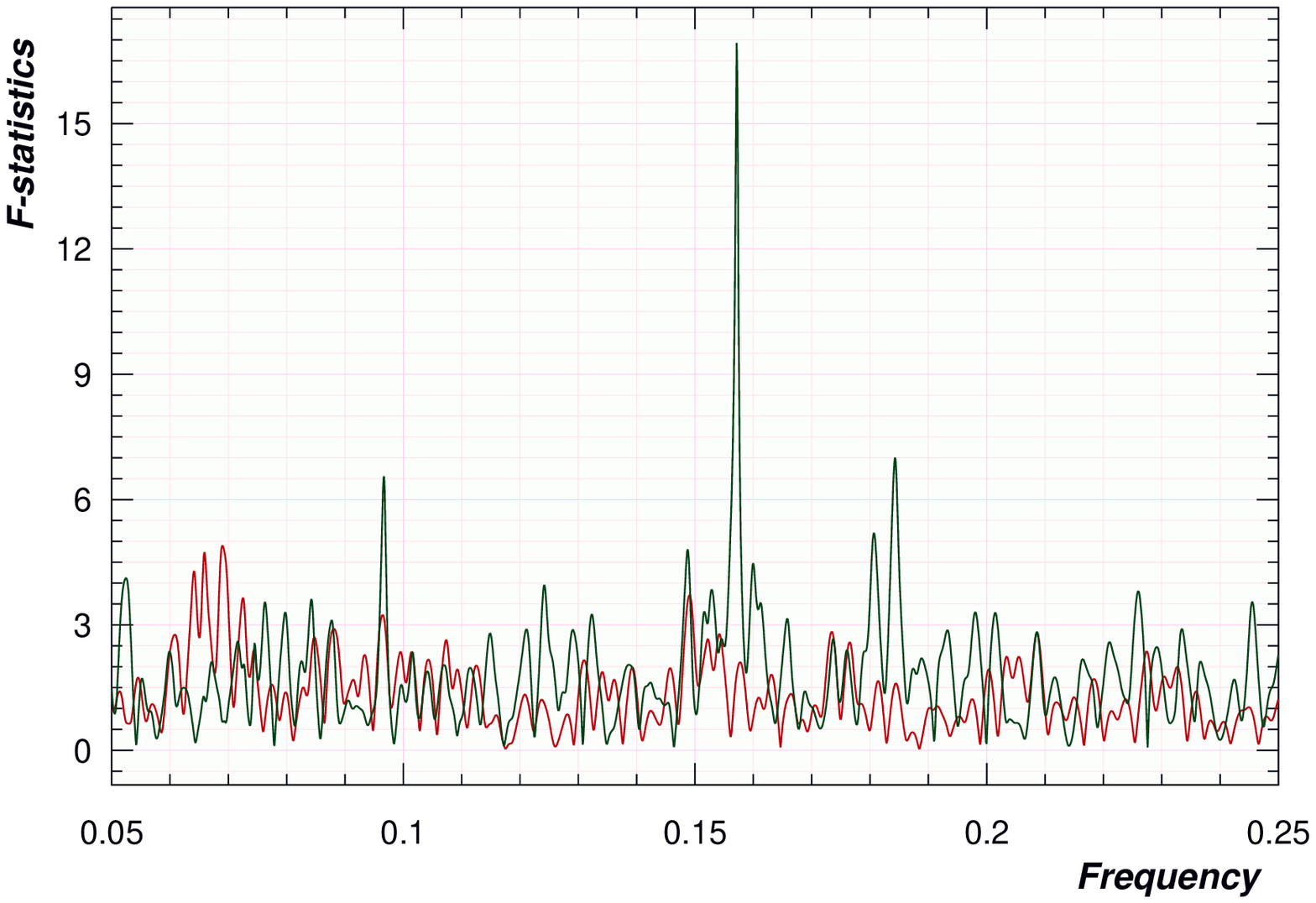}
\caption{
{\it Left panel}: F-statistics frequency spectrum (in d$^{-1}$) for the longitudinal magnetic field measurements of  
CPD\,$-$57$^{\circ}$3509 using the hydrogen lines. {\it Right panel}: F-statistics frequency spectrum using 
the entire spectrum for the measurements. The window function is indicated by the red color. 
}
\label{fig:period}
\end{figure*}

The results of our frequency analysis based on the longitudinal magnetic field measurements 
presented in Table~\ref{tab:log_meas} and 
performed using a non-linear least squares fit to the multiple harmonics utilizing the Levenberg-Marquardt
method \citep{Press1992} 
are presented in Fig.~\ref{fig:period}.
To detect the most probable period, we calculated the frequency spectrum and for each trial
frequency we performed a statistical F-test of the null hypothesis for the absence of periodicity 
\citep{Seber1977}. The resulting F-statistics can be thought of as the total sum including covariances of the ratio 
of harmonic amplitudes to their standard deviations, i.e.\ a signal-to-noise ratio.
The highest peak in the frequency spectrum for the hydrogen lines not coinciding with the window function
corresponds to a period of $6.36093\pm0.00026$\,d and that for the measurements of the entire spectrum
corresponds to a period of $6.36255\pm0.00026$\,d.
We note that taking into account the precision 
of the period determination, the difference between the two periods is not significant.
On the other hand, 
as the achieved magnetic field measurement accuracy is higher for the set using all lines, the rotation 
period of 6.36255\,d identified from these  measurements is expected to 
be more reliable and is preferred in the following discussion on the surface element distribution in 
Sect.~\ref{sect:doppler}. 

\begin{figure*}
\centering
\includegraphics[width=0.45\textwidth]{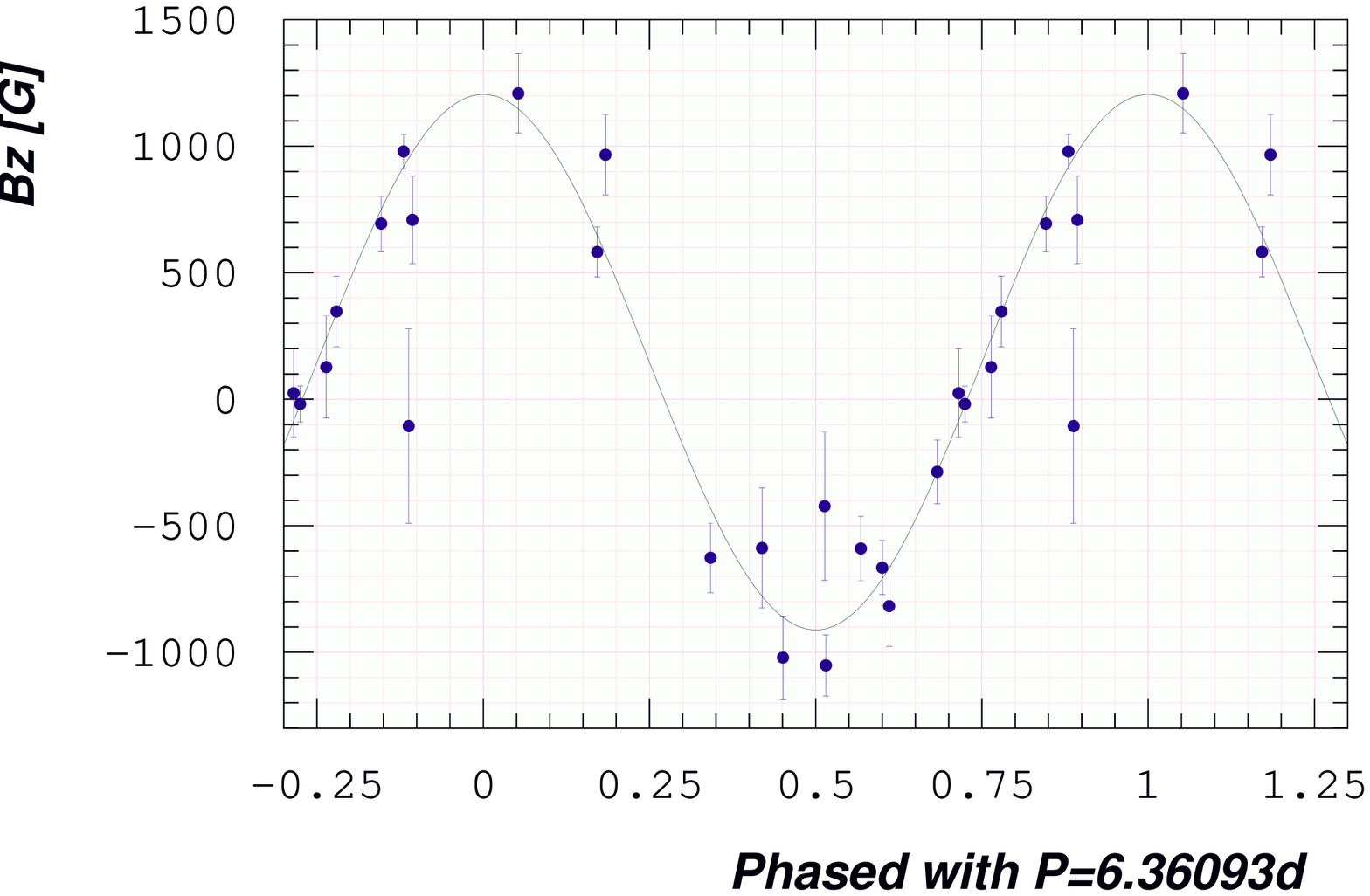}
\includegraphics[width=0.45\textwidth]{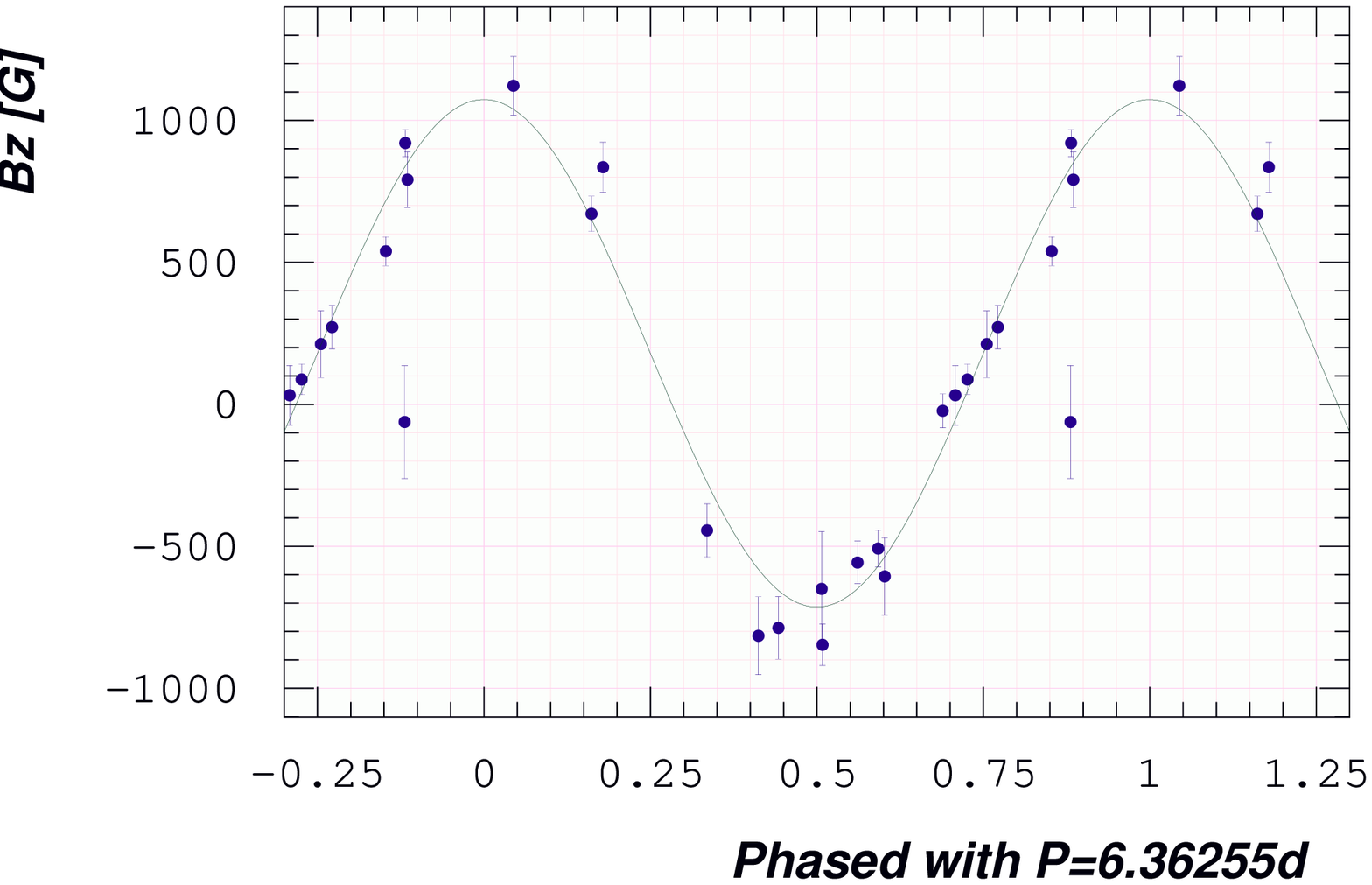}
\caption{
{\it Left panel}: Longitudinal magnetic field variation of CPD\,$-$57$^{\circ}$3509 measured using the hydrogen lines
phased with the 6.36093\,d period.
The solid line represents a fit to the data with a mean value for the magnetic field of
$\overline{\left<B_{\rm z}\right>} = 147\pm52$\,G, and an amplitude of
$A_{\left<B_{\rm z}\right>} = 1058\pm64$\,G.
For the presented fit, we assume a zero phase corresponding to a positive field extremum at $MJD56983.5063\pm0.0590$.
{\it Right panel}: Longitudinal magnetic field variation of CPD\,$-$57$^{\circ}$3509 measured using the entire 
spectrum phased with the 6.36255\,d period.
The solid line represents a fit to the data with a mean value for the magnetic field of
$\overline{\left<B_{\rm z}\right>} = 180\pm47$\,G and an amplitude of
$A_{\left<B_{\rm z}\right>} = 894\pm57$\,G.
For this fit, we assume a zero phase corresponding to a positive field extremum at $MJD56983.5396\pm0.0646$.
Please note that values below 0 and above 1 are repetitions
and plotted to visualize the transition at the intervall borders.
}
\label{fig:rot}
\end{figure*}

In Fig.~\ref{fig:rot}, we present all measurements, those using the entire spectrum and those using only the
hydrogen lines, phased with the corresponding rotation periods  and the best sinusoidal fits calculated for these 
measurements.
As already mentioned in the caption of Table~\ref{tab:log_meas},
the deviating point close to the phases 0.881, respective 0.888, is caused by the unfavourable weather during the observations at that phase.

\begin{figure*}
  \centering
  \includegraphics[width=0.35\textwidth]{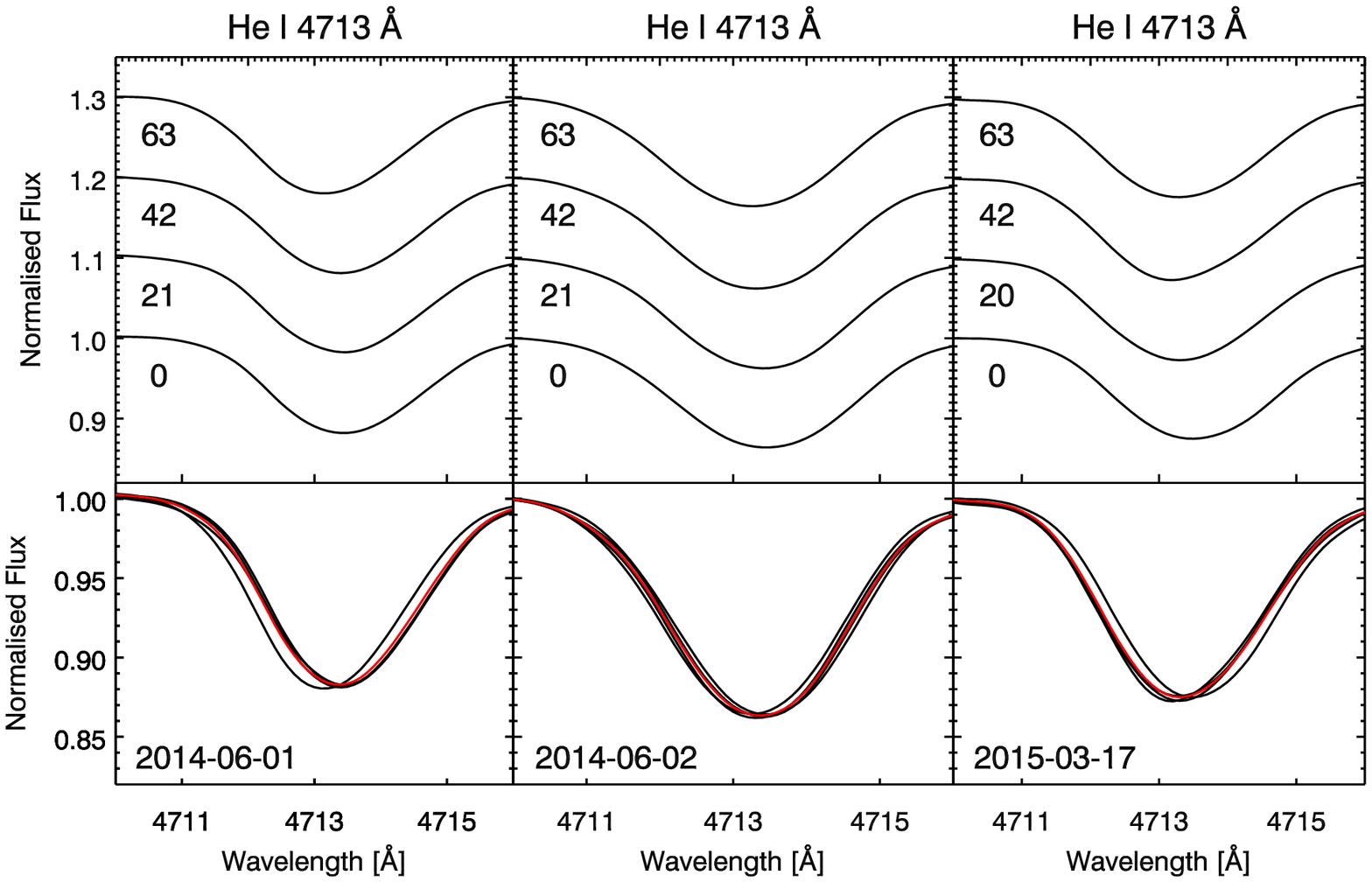}
\includegraphics[width=0.62\textwidth]{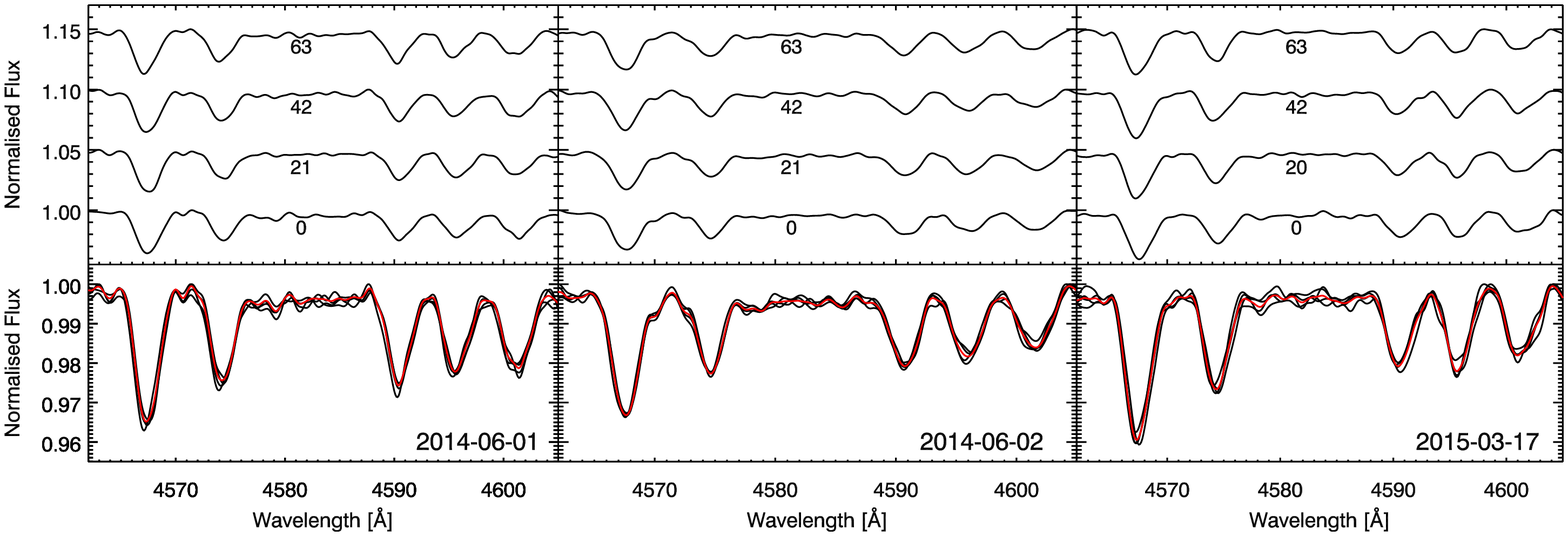}
\caption{
{\it Left panel}: The behaviour of \ion{He}{i}~4713 in the FORS\,2  spectra
in each individual subexposure belonging to observations on three different epochs. For each epoch, in the upper row,
we present the line profiles shifted in vertical direction for best visibility.
The time difference (in minutes) between the subexposure and the beginning of the
observations is given close to each profile. The lower row shows all profiles
overplotted. The average profile is indicated by the red line.
{\it Right panel}: The same as in the left panel, but for the metallic lines in the spectral region 4562--4605\,\AA{}. 
  }
  \label{fig:puls_c}
\end{figure*}

  Since CPD\,$-$57$^{\circ}$3509 already finished half of its main-sequence lifetime \citep{Przybilla2016} 
and is already passing through the $\beta$~Cep instability strip,
we checked  the stability 
of the Stokes $I$ spectral lines over the full sequences of sub-exposures obtained on a time scale of
tens of minutes. Along with different radial 
velocity shifts of lines  belonging to different elements, we also detect 
distinct changes in line profiles taking place on time-scales corresponding to the duration of the
sub-exposure sequences in the individual observations. In Fig.~\ref{fig:puls_c},
we present the behaviour of the line profiles in individual spectral lines.
The time difference between subexposures accounts for about 20\,min.
However, with the current data we cannot identify the periodicity of the detected variability,
which is probably caused by the presence of $\beta$~Cep-like pulsations.
Thus, future observations should focus on the 
careful search for periodicity and on the identification  of the pulsation  modes.

\begin{figure}
\centering
\includegraphics[width=.45\textwidth]{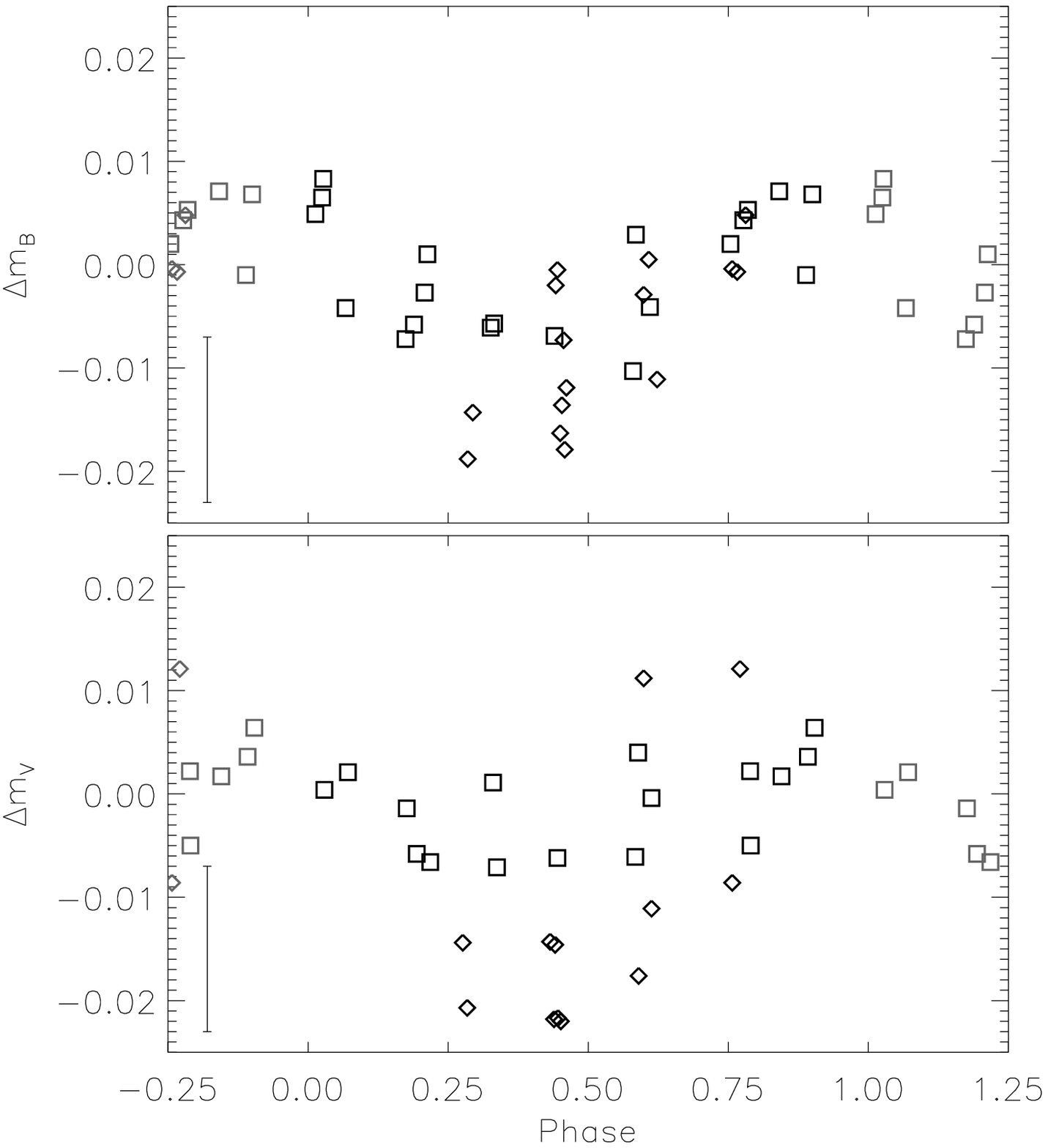}
\caption{
Differential photometry of CPD\,$-57^{\circ}$3509 in the
$B$ (upper panel) and $V$ (lower panel) bands phased with the period from the
magnetic field analysis using the hydrogen lines.
Data from 2014 are displayed as boxes, data from 2015 as diamonds.
A (conservative) error bar for the individual measurements is shown at the bottom left in each panel.
Please note that values below 0 and above 1 are repetitions
and plotted to visualize the transition at the intervall borders.
}
\label{fig:photometry}
\end{figure} 

The periodicity derived from spectropolarimetry (and the corresponding spottiness,
see below) plus the tentative $\beta$\,Cep-like pulsations are expected to be detectable
in the photometric data. Our differential photometry data in
$B$ and $V$ are displayed in Fig.~\ref{fig:photometry}, phased
according to the period obtained from the magnetic analysis using
only the hydrogen lines -- the behaviour is very similar for the case of the phasing
based on all lines. A small periodic variability in
line with the changes of the magnetic field strength is clearly seen.
The dispersion of the higher-cadence data of 2015 (taken within $\sim$2/3 of the
rotation period) could be interpreted in favour of the presence of
$\beta$\,Cep pulsations, but this is close to the detection limit and
requires dedicated follow-up observations for confirmation.
Finally, we want to note a difference in the mean $B$ and $V$
magnitudes from the present work and that of
\citet{Baume2003}, which are 0\fm07 and 0\fm06 fainter than
our values. As the comparison stars used for the differential
photometry show only an rms of 0\fm02 between both studies (data taken
$\sim$20 years apart), this may imply some long-term variability for
CPD\,$-57^{\circ}$3509 as well.

\section{Doppler imaging using UVES observations}
\label{sect:doppler}

As we mentioned above, the rotational phase coverage and S/N of the UVES observations is quite good and 
enables mapping the chemical element patterns, in particular of the most strongly varying 
elements He and Si. 
The phase coverage of the observations is not optimal, though, and four observations have been obtained 
close in phase. Still, the largest phase gap is 0.22, and occurs between phases 0.79 and 0.01. 
Other areas on the stellar surface are well covered, and the phase gaps are less than 0.15,
posing no problems for the Doppler imaging technique.  
Tests show that phase gaps as large as about 0.28 do not significantly affect the reconstruction 
of the features \citep{Rice2000} if the S/N is good. Commonly seen problems caused by non-perfect
phase coverage are the blurring of features and not being able to recover small spots (see e.g., 
\citealt{Collier1994}). Chemical spots are large structures 
and therefore can easily be mapped with the phase coverage of the current observations. Naturally, 
one should be careful when interpreting the results, especially small features, in the location of the 
largest phase gap. In Sect.~\ref{sect:maps}, the discussion includes the possible impact of the phase gap
on the obtained maps.

\begin{figure*}
  \centering
  \includegraphics[width=0.90\textwidth]{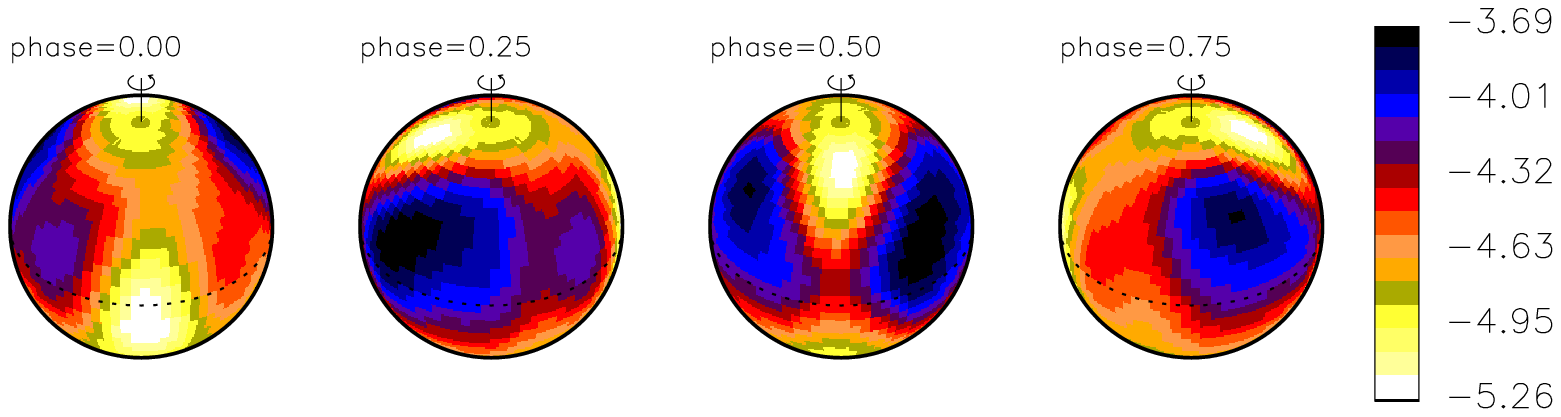}
  \includegraphics[width=0.90\textwidth]{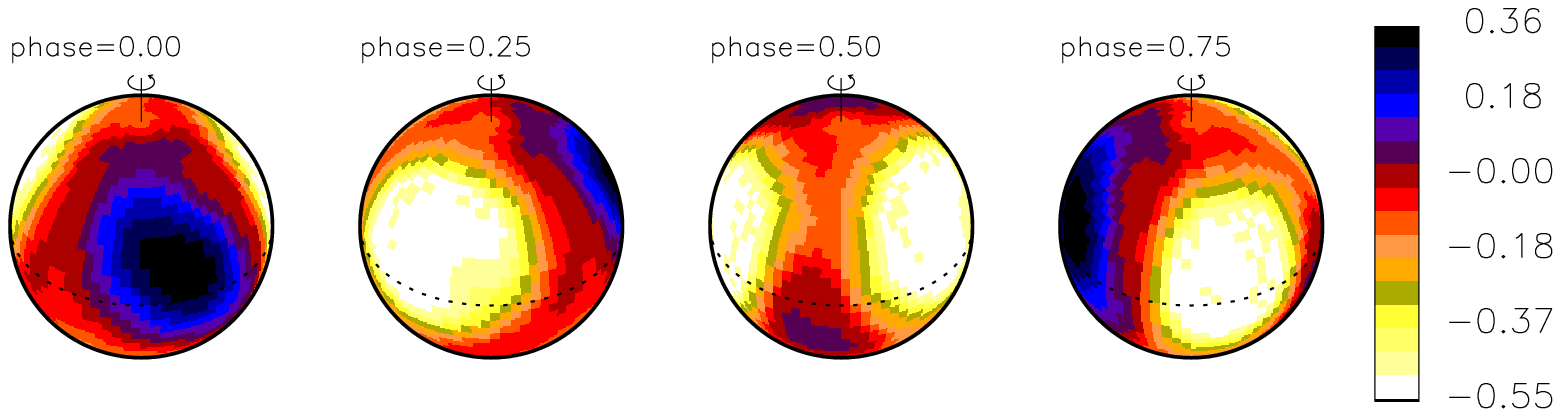}
  \caption{Chemical abundance maps of CPD\,$-57^{\circ}$3509 for \ion{Si}{iii} (top) and \ion{He}{i} 4713 (bottom). 
The \ion{Si}{iii} map has been obtained simultaneously from three lines:  \ion{Si}{iii} 4553, 
\ion{Si}{iii} 4568, and \ion{Si}{iii} 4575. The surface distribution of both elements is shown for 
four different phases that are 0.25 apart. The abundance is given with respect to the total number density of 
atoms and ions.}
  \label{fig:maps}
\end{figure*}

Since observations with HARPSpol are usually carried out in visitor mode, we 
do not have high-resolution
spectropolarimetric observations of  CPD\,$-57^{\circ}$3509 on our disposal. The magnetic field strength is
however only a few kG, and as we show below, it is possible to use the Doppler imaging technique 
to obtain an approximate elemental distribution without taking into account the impact of the magnetic field on the 
line profiles.
The He mapping is carried out using \ion{He}{i} 4713. For \ion{Si}{iii},
three lines located close to each other, 4552.622, 4567.840, and 4574.757\,{\AA}, were used simultaneously. 
For the phase calculation we employed the rotation period of $6.36255\pm0.00026$\,d
identified in our magnetic field measurements using the entire spectrum.
The surface abundance maps were obtained with the INVERS7PD inversion code, which was originally developed 
by \citet{pisk90} and modified by \citet{hack}. INVERS7PD compares the observations to a grid 
of synthetic local line-profiles, which were calculated using the SPECTRUM spectral synthesis code 
\citep{SPECTRUM} and ATLAS9 stellar atmospheres by 
\citet{Kurucz}. The \ion{He}{i} line properties are hard coded into SPECTRUM, and the \ion{Si}{iii} line 
parameters were originally obtained from VALD \citep{VALD1,VALD2}. 
When using the Si abundance determined by \citet{Przybilla2016}, the three \ion{Si}{iii} lines were on average 
well fitted, but the 4552.622\,{\AA} line was slightly too weak, while the 4574.7570\,{\AA} line 
appeared slightly too strong. To 
minimise the influence of the errors in the atomic data, we followed the standard procedure and changed the 
oscillator strengths of the lines to improve the overall fit 
(see e.g., \citealt{Makaganiuk2011,Korhonen2013}). The changes needed were small and allowed for 
fitting all three \ion{Si}{iii} lines simultaneously. Table~\ref{tab:Si} gives the parameters used for these lines. 

\begin{table}
  \caption[]{\ion{Si}{iii} $\log gf$ values used in this work.}
  \centering
  \label{tab:Si}
  \begin{tabular}{crr}
    \hline
    \hline
    \noalign{\smallskip}
\multicolumn{1}{c}{Wavelength} &
\multicolumn{1}{c}{$\log gf$} &
\multicolumn{1}{c}{$\log gf$} \\
\multicolumn{1}{c}{[\AA{}]} &
\multicolumn{1}{c}{VALD} &
\multicolumn{1}{c}{this work} \\
    \noalign{\smallskip}
    \hline
    \noalign{\smallskip}
    4552.6220 &  0.181 &  0.275 \\
    4567.8400 & -0.039 & -0.039 \\
    4574.7570 & -0.509 & -0.709 \\
    \noalign{\smallskip}
    \hline
  \end{tabular}
\end{table}

For the model calculations, 20 limb angles were used together with the stellar parameters 
determined by \citet{Przybilla2016}: $T_{\rm eff}=23\,750$\,K, $\log g = 4.0$, and micro- and macroturbulence of 2.0 
and 10.0 ${\rm km\,s}^{-1}$, respectively.
From the inversions, we obtained a best fit inclination angle of $58\pm10^{\circ}$.
Notably, the Doppler imaging technique is very sensitive to the $v \sin i$ and 
the inclination angle values in the sense that no converging solution can be found if these values are 
under- or overestimated. 

\begin{figure*}
  \centering
  \includegraphics[width=0.68\textwidth]{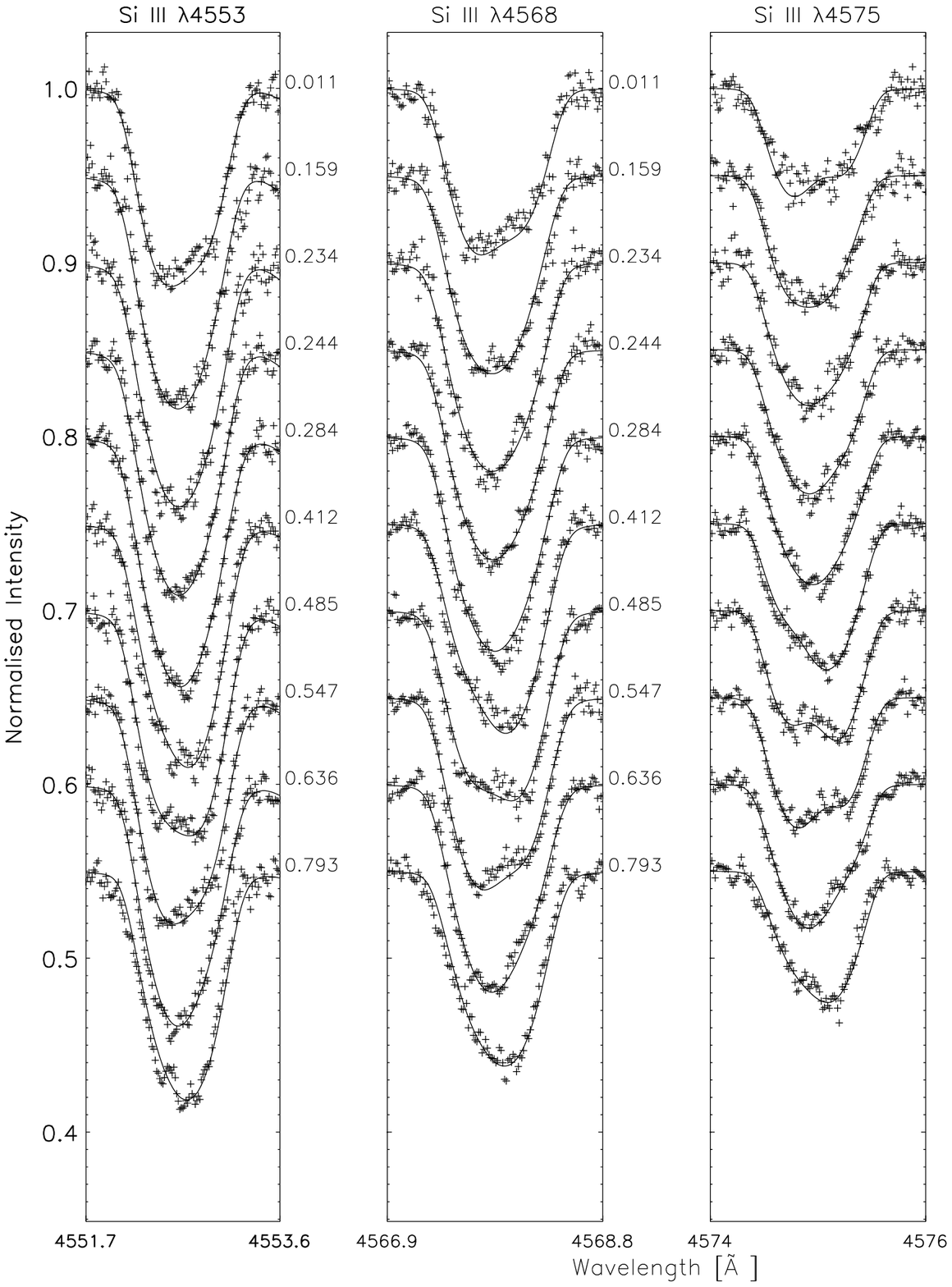}
  \includegraphics[width=0.225\textwidth]{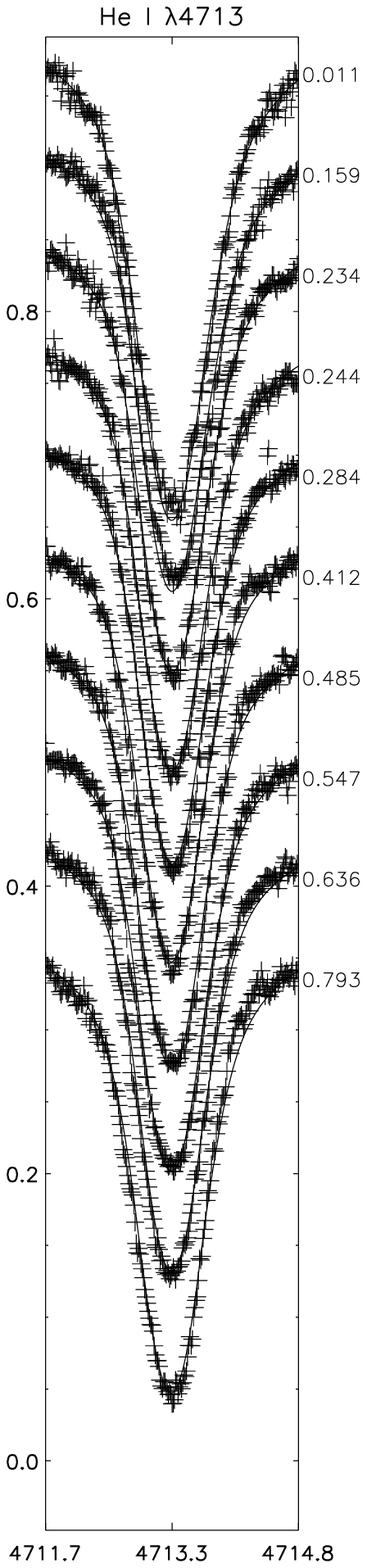}
  \caption{Spectroscopic observations (plus signs) together with the model fit from the inversions (solid line). 
The fits are shown for all the lines used in the inversions (from left to right): \ion{Si}{iii} 4553, 
\ion{Si}{iii} 4568, \ion{Si}{iii} 4575, and \ion{He}{i} 4713. Line profiles are shifted in vertical 
direction for better visibility.}
  \label{fig:spectra}
\end{figure*}

\citet{Przybilla2016} estimated the $v\sin i$ value of CPD\,$-57^{\circ}$3509 to be $35\pm2\,{\rm km\,s}^{-1}$. In the inversions, the 
best fit from the \ion{Si}{iii} lines was obtained for $v\sin i = 34.5\,{\rm km\,s}^{-1}$, in excellent agreement
with the results of \citet{Przybilla2016}. On the other hand, for the \ion{He}{i} line, the best fit was obtained for 
$v\sin i = 26.5\,{\rm km\,s}^{-1}$, which is lower than the other estimates. There are several 
possible reasons for this discrepancy. First, weaker metal lines are more sensitive to $v\sin i$ values 
and broad \ion{He}{i} lines dominated by Stark effect are usually avoided in the 
determination of rotation rates. 
\ion{Si}{iii} and \ion{He}{i} lines also have different Land\'e factors.
Further, the star is likely a $\beta$\,Cep--like pulsator. Depending on the type of pulsations, this 
can have more impact on the \ion{Si}{iii} lines rather than on the \ion{He}{i} lines, causing stronger line
broadening in the \ion{Si}{iii} lines. We also note that similar discrepancies in the determination of $v\sin i$ 
values for different elements were mentioned in other works, which were using imaging methods 
(e.g.\ \citealt{Yakunin2015}).
We also remark that SPECTRUM is verified to work at the spectral type 
range B to mid-M, making CPD\,$-57^{\circ}$3509 close to the limit of the code's capabilities. Furthermore, SPECTRUM does 
not use NLTE in the spectral synthesis, and \ion{He}{i} is more sensitive to departures from NLTE 
than \ion{Si}{iii}. For these reasons, the absolute abundance scales in the obtained maps are not necessarily 
precise. Still, this will not affect the relative abundances of the spots, nor their locations. 

\subsection{Magnetic field geometry}
\label{sect:geom}

The simplest model for a magnetic field geometry is based on the assumption that the studied stars 
are oblique dipole rotators,
i.e., their magnetic field can be approximated by a dipole with its magnetic axis 
inclined with respect to the rotation axis.

From the variation of the phase curve for the
field measurements with a mean of 
$\overline{\left< B_{\rm z}\right>} = 180\pm 47$\,G and an amplitude of
$A_{\left< B_{\rm z}\right>} = 894 \pm 57$\,G, we calculate
$\left< B_{\rm z} \right>^{\rm min}= -714\pm74$\,G and
$\left< B_{\rm z} \right>^{\rm max}=1074\pm74$\,G.
Using the definition by \citet{Preston1967}

\begin{equation}
r = \frac{\left< B_{\rm z}\right>^{\rm min}}{\left< B_{\rm z}\right>^{\rm max}}
  = \frac{\cos \beta \cos i - \sin \beta \sin i}{\cos \beta \cos i + \sin \beta
\sin i},
\end{equation}

\noindent
we find 
$r= -0.665\pm0.075$ and finally following

\begin{equation}
\beta =  \arctan \left[ \left( \frac{1-r}{1+r} \right) \cot i \right],
\end{equation}

\noindent
and employing $i=58\pm10^{\circ}$ obtained from our 
Doppler imaging inversions, we calculate a magnetic obliquity angle $\beta=72\pm8^{\circ}$. 
Assuming a limb-darkening coefficient $u=0.3$, typical for stars with $T_{\rm eff}=23\,750$\,K \citep{Przybilla2016},
we estimate a dipole strength of $3.91\pm0.36$\,kG using the model by \citet{Stibbs1950}, 
as formulated by \citet{Preston1967}:
\begin{equation}
B_{\rm d} = \left< B_{\rm z}\right>^{\rm max}  \left( \frac{15+u}{20(3-u)} (\cos \beta \cos i + \sin \beta \sin i) \right)^{-1}.
\end{equation}

Using the parameters of the sinusoidal fit to the values resulting from only the hydrogen lines
($\overline{\left<B_{\rm z}\right>} = 147\pm52$\,G and $A_{\left<B_{\rm z}\right>} = 1058\pm64$\,G),
we obtain slightly different values for the magnetic field model:
$r= -0.757\pm0.077$,
$\beta=78\pm6^{\circ}$, and 
$B_{\rm d} = 4.51\pm0.45$\,kG. Given the size of the errors of the dipole strength determination,
the difference in the derived values of the dipole strengths is not significant.

\subsection{Abundance maps and comparison to the magnetic pole location}
\label{sect:maps}

The \ion{He}{i} and \ion{Si}{iii} distributions are shown in Fig.~\ref{fig:maps}, and the model fits to the 
observations are given in Fig.~\ref{fig:spectra}. The locations of the chemical spots are very similar in both maps, 
but the behaviour is opposite. The main concentration of \ion{He}{i} occurs at phase 0.0, and a weaker 
concentration around the phase 0.5. The \ion{He}{i} underabundance spots are located at phases 0.4 and 
0.7. In contrast, the main concentrations of \ion{Si}{iii} occur around the phases 0.4 and 0.7, and 
the underabundance 
spots are located around the phases 0.0 and 0.5.
In \ion{Si}{iii}, also a third smaller overabundance spot is 
seen around the phase 0.1.
Some indication of a corresponding spot could also be seen around phase 0.9, 
but due to the phase gap in the observations at phases 0.78--0.01,
we cannot verify the reality of this potential fourth \ion{Si}{iii} spot.

\begin{figure}
  \centering
  \includegraphics[width=0.49\textwidth]{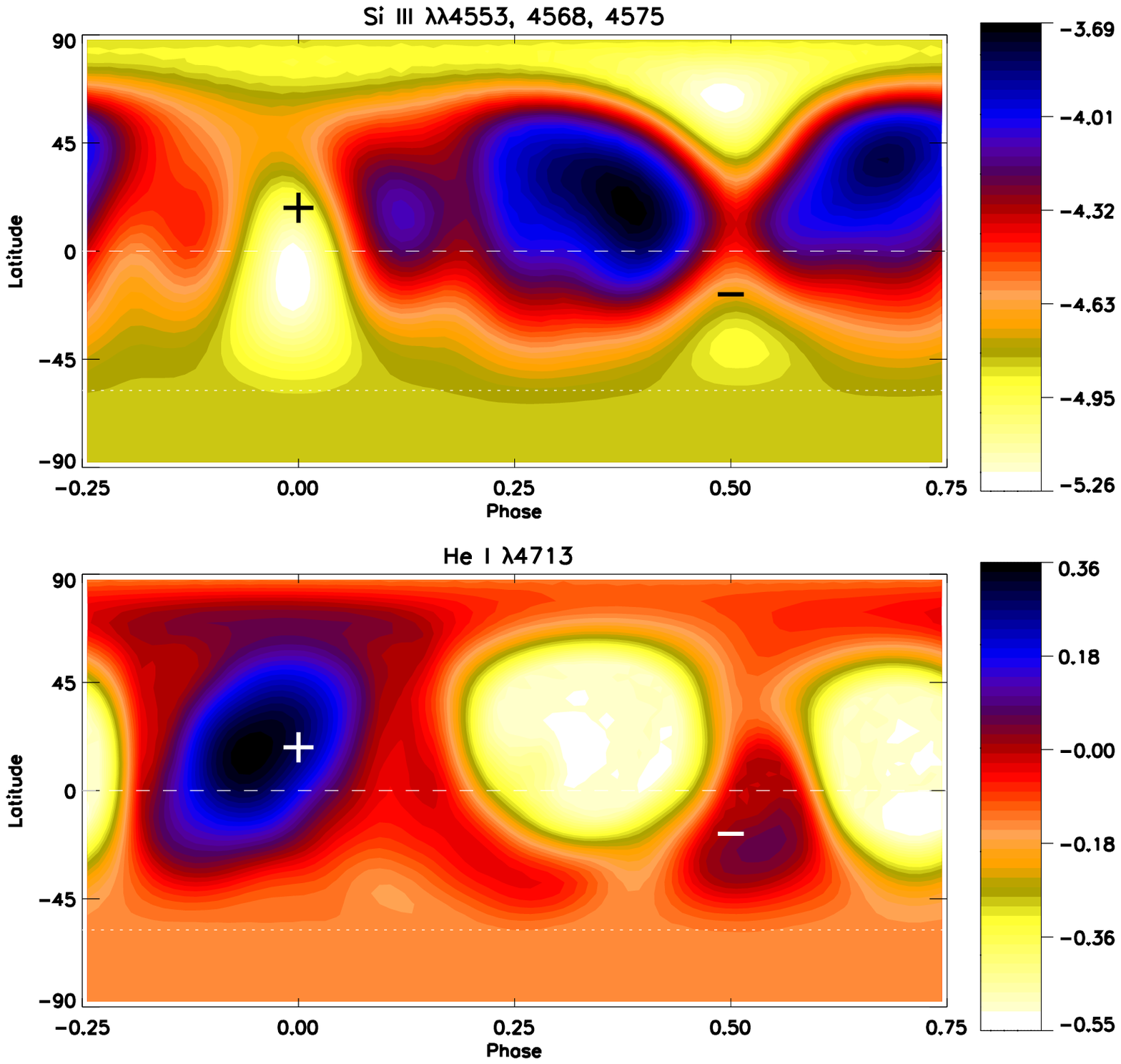}
  \caption{Mercator abundance maps plotted together with the location of the magnetic poles. As before, \ion{Si}{iii} 
is plotted at the top and \ion{He}{i} at the bottom. The plus-sign denotes the location of the positive magnetic 
pole, and the negative pole is given by the minus-sign. The dashed line shows the stellar rotational equator, and 
the dotted line gives the limit below which the star is not seen due to its inclination. Note that the phases go 
from -0.25 to 0.75 to better show the area around the magnetic poles.}
  \label{fig:maps_mg}
\end{figure}

Figure~\ref{fig:maps_mg} shows the locations of the magnetic poles together with the abundance maps. The 
\ion{He}{i} overabundance spots clearly occur around the magnetic poles, with a stronger concentration at 
the positive pole and a weaker one around the negative pole. The locations do not coincide completely in phase with the 
magnetic poles, and the overabundance spots seem to be shifted some 0.1 in phase away from the poles. On the 
other hand, \ion{Si}{iii} shows underabundance around the magnetic poles, and the main overabundance spots 
fall closer to the magnetic equator. Interestingly, it seems that the \ion{Si}{iii} concentrations occur 
somewhat shifted towards the negative pole, not at the magnetic equator itself. Similarly, the 
underabundance spots of \ion{He}{i} are located closer to the negative magnetic pole.

The presented Doppler maps for Si and He support the dipole-dominated magnetic topology  of 
CPD\,$-57^{\circ}$3509.
On the other hand, the presence of a third Si spot detected around the phase 0.1 and  
considerable abundance and field strength differences between the two magnetic poles suggest the contribution
of a non-negligible higher-order magnetic multipole.
Previous studies of upper main-sequence stars showed that inhomogeneous chemical abundance distributions
are only observed  on  the  surface  of  magnetic  chemically  peculiar Ap and Bp stars with large-scale 
organised magnetic fields. In these stars, the abundance distribution of certain elements 
is non-uniform and non-symmetric with respect to the rotation axis. The majority of studies of Ap and Bp stars  
have revealed a kind of symmetry between the topology of the magnetic field and the element distribution
(see e.g.\ \citealt{Rice1997,Yakunin2015,Hubrig2014b}). Thus, the structure of the magnetic field 
can be studied by producing the 
surface element maps and measuring the magnetic
field using spectral lines of inhomogeneously distributed elements separately. However, due to the low 
resolution of our 
FORS\,2 spectra, we are not able to study the detailed surface magnetic field distribution.
Furthermore, only the availability of high-resolution spectra in all four Stokes parameters
would allow us to obtain
self-consistent mapping of spots and magnetic fields by means of Zeeman Doppler imaging (ZDI; 
e.g.\ \citealt{Brown1991}).

\section{Discussion}
\label{sect:disc}

Our spectropolarimetric monitoring of CPD\,$-57^{\circ}$3509 using FORS\,2 at the VLT shows 
the presence of an approximately dipolar magnetic field with a polar
strength of 4\,kG, reversing over the rotation period of 6.36\,d. Using the Doppler imaging technique,
we were able to constrain the inclination of the rotation axis to the line of sight, $i=58\pm10^{\circ}$, 
and estimate the obliquity of the magnetic axis, $\beta=72\pm8^{\circ}$.
In the past it was considered neither theoretically nor observationally how the 
magnetic field geometry in massive B-type stars evolves across the main sequence. Thus, the analysis of
the magnetic field configuration of CPD\,$-57^{\circ}$3509 is of special interest as it
is one of the most evolved He-strong stars currently known. \citet{Hubrig2007} studied the 
evolution of the magnetic field geometry in late B-type stars with masses $\leq5\,M_\odot$ with
accurate Hipparcos parallaxes and definitely determined longitudinal magnetic fields. 
They found that 
the distribution of relative ages peaks at the ZAMS, with two secondary lower peaks at the relative ages 
around 60\% and 80\%. The strongest magnetic fields were found in younger stars in terms of the elapsed fraction
of their main-sequence life. Further, rotation periods of late-B type stars slightly increase with age, which is 
consistent with
the assumption of conservation of angular momentum during their life on the main sequence, without any hint of a
braking mechanism (see also \citealt{North1984,North1985}). 
The fact that the strongest magnetic fields are only observed close to the ZAMS can be
interpreted as a magnetic field decay in stars at advanced ages. As for the magnetic field geometry,
the authors detected a strong hint for an increase of obliquity $\beta$ with elapsed time on the main sequence.
Moreover, 21\% out of the studied 33 stars have magnetic phase curves fitted by a
double wave, indicating that the magnetic topology in late-B type stars is frequently more complex than
just a single dipole. 
Obviously, a comparison of the evolution of magnetic field geometries between the 
higher-mass He-strong stars and the lower-mass magnetic B-type stars is urgently needed to constrain the mechanism 
of the magnetic field generation in such stars.

There is also a large dissimilarity between the lower mass magnetic Ap stars and the He-strong stars in respect to 
the orientation of their magnetic axes and the period lengths. 
The results of modeling a small sample of Ap stars by \citet{Landstreet2000} implied that 
stars with small obliquity values $\beta$ of the model magnetic axis to the rotation axis,
of  the  order of  $20^{\circ}$, have periods longer than 25~days.
However, recent studies of the magnetic field geometries in He-strong stars do not confirm this trend: a number of 
fast rotating He-strong stars with periods below 2\,d show low obliquities of their magnetic axes 
(e.g.\ \citealt{Grunhut2012,Sikora2015,Hubrig2017}).

The study of the variability of the Si and He lines showed the presence of significant chemical
abundance variations across the stellar photosphere. The location of the chemical spots is roughly correlated 
with the topology of the magnetic field, where the main concentration of He is observed in the vicinity 
of the 
positive magnetic pole and Si underabundant spots around the poles.
Only three mapping studies were devoted to He-strong stars in the past,
all of them using ZDI \citep{Yakunin2015,Oksala2015,Kochukhov2011}. 
\citet{Yakunin2015} studied the He and O abundance distribution on the surface of the He-strong 
star HD\,184927.
Similar to our DI result, the He abundance was the highest in the vicinity of the strongest 
magnetic pole which in that case was the pole of positive polarity. \citet{Oksala2015} studied 
the distribution of He, C, Si, and Fe on the surface of $\sigma$\,Ori\,E.
A large overabundant He spot was found to appear at the rotation phase 0.8, but the location of the 
spot was not correlated with the position of the magnetic poles. The minimum abundances of 
C, Si, and Fe were found at the same phase where He showed the largest abundance.
Finally, \citet{Kochukhov2011} studied the distribution of He on the surface of HD\,37776 
and concluded that the He concentration is at a maximum in the regions of maximum radial field.
As only very few He-strong stars were studied with ZDI, future high-resolution, high signal-to-noise 
spectropolarimetric observations are needed to obtain a more complete picture on how
the distribution of surface chemical spots is related to the magnetic field topology.
We note that CPD\,$-57^{\circ}$3509 has a relatively low projected rotational velocity, and due to the presence of 
the kG magnetic field and the distinct inhomogeneous element abundance distribution, 
it can serve as an excellent 
laboratory to study various atmospheric effects that interact with the magnetic field.
Further, the discovered pulsational variability on the time scale of tens of minutes has to be confirmed by
future high-resolution spectroscopic time series.

\section*{Acknowledgments}
\label{sect:ackn}

Based on observations made with ESO Telescopes at the La Silla Paranal Observatory under programmes 094.D-0355
and 191.D-0255.
The observations on Cerro Armazones are supported by the Nordrhein-Westf\"{a}lische
Akademie der Wissenschaften und der K\"{u}nste in the framework of the
academy program of the Federal Republic of Germany and the state
Nordrhein-Westfalen. AK acknowledges financial support from RFBR grant 16-02-00604A.

\appendix

\section{The membership of CPD$-$57$^{\circ}$3509 in NGC\,3293}
\label{sect:appA}

The star CPD$-$57$^{\circ}$3509 is located only 66\,arcsec from the cluster
centre of NGC\,3293, whereas the angular radii
of the cluster core, of the central part of the cluster, and of the whole
cluster were determined as 72\,arcsec, 306\,arcsec, and 486\,arcsec, 
respectively
\citep{kharchenko13}.
Although this relatively bright star is listed in the Tycho catalogue
\citep{esa97}, it has no proper motion measurement
in the Tycho-2 catalogue \citep{hog00}. Therefore, it was
not included in the catalogue of stars in open cluster areas of
\citet{kharchenko04}, where membership probabilities
are given for 520 open clusters of the survey of
\citet{kharchenko05} based on proper motions and
optical photometry. In the new Milky Way Star Clusters (MWSC) survey by
\citet{kharchenko13}, based on proper motions from
the PPMXL catalogue \citep{roeser10} and
near-infrared photometry from the Two Micron All Sky Survey (2MASS;
\citealt{skrutskie06}), CPD$-$57$^{\circ}$3509 is however
listed with a 95\% membership probability from proper motion and 99\%
from its $JK_s$ photometry.
\citet{dias14} also determined a high (98\%) membership
probability based only on proper motions from the fourth US Naval
Observatory CCD Astrograph Catalog (UCAC4; \citealt{zacharias13}).

One should however note that the proper motion errors of the catalogues
used for the above mentioned cluster membership studies are rather large.
In case of CPD$-$57$^{\circ}$3509 they are $\pm$2.5\,mas/yr in the PPMXL
and between $\pm$2.2\,mas/yr and $\pm$4.0\,mas/yr in the UCAC4 catalogue. 
A significant
improvement of PPMXL proper motions was recently achieved with the
Hot Stuff for One Year (HSOY; \citealt{altmann17}) catalogue,
which represents a new reduction of the PPMXL including Gaia DR1 data
\citep{gaia16a,gaia16b,lindegren16}.
The new UCAC5 catalogue
\citep{zacharias17} presents very accurate proper
motions combining the UCAC positions with those from Gaia DR1.
Because CPD$-$57$^{\circ}$3509
was not in the Tycho-2 catalogue, it was not included in the Tycho-Gaia
Astrometric Solution (TGAS) of Gaia DR1. However, its HSOY proper motion,
$\mu_{\alpha}\cos{\delta}=-5.9\pm1.1$\,mas/yr, 
$\mu_{\delta}=+3.5\pm1.1$\,mas/yr,
and its UCAC5 proper motion, $-6.1\pm0.9$\,mas/yr, $+3.2\pm0.9$\,mas/yr,
are in very good agreement with the mean TGAS proper motion of four
of the five 1$\sigma$ members from \citet{kharchenko04}
that can be found in TGAS: $-6.7\pm1.2$\,mas/yr, $+3.7\pm0.3$\,mas/yr. Their
standard deviation in $\mu_{\alpha}\cos{\delta}$ is still about three times
larger than in $\mu_{\delta}$ but was even much larger before we excluded
one of the five stars as an outlier.

We do not consider the TGAS parallaxes of the
few 1$\sigma$ members from \citet{kharchenko04},
as they show a large spread, and since CPD$-$57$^{\circ}$3509 is not
included in the TGAS. The very large standard deviation of the
parallaxes of the five 1$\sigma$ members ($\pm$0.77\,mas) reduces to
$\pm$0.19\,mas, if we again exclude the same one outlier. However, the
generally assumed additional systematic error of $\pm0.3$\,mas in TGAS
parallaxes \citep{gaia16b} leads to a
$\>60$\% relative distance uncertainty of a cluster at about 2\,kpc distance.
According to \citet{kharchenko05,kharchenko13},
the distance to NGC\,3293 is about 2400\,pc.

Concerning the radial velocity of CPD$-$57$^{\circ}$3509, there is
one measurement, $-16$\,km\,s$^{-1}$ given without an error estimate
\citep{Evans2005}, and slight variability $-16$...$-20$\,km\,s$^{-1}$
measured by \citet{Przybilla2016} due to the presence of spots.
This is in reasonably good agreement with the mean cluster radial velocities
determined by \citet{kharchenko05,kharchenko13} and \citet{Evans2005},
of $-12.3\pm2.3$\,km\,s$^{-1}$, $-11.2\pm2.1$\,km\,s$^{-1}$,
and $-12\pm5$\,km\,s$^{-1}$,
respectively.

Thus the membership of CPD$-$57$^{\circ}$3509 in the cluster NGC\,3293 is
based on the star's projection to the cluster core, and its available proper
motion, photometry, and radial velocity. Gaia DR2, expected for
April 2018, will not only provide even more accurate proper motion membership
probabilities but also enable a membership study using individual stellar
parallaxes of many more stars in the cluster area.

\section{Photometric data}
\label{sect:appB}

%
%
%
%
%
%
%
%
%
%
%
%
%
%
%
%
%
%
%
%
%
%
%
%
%
%
%
%
%
%
%
%
%
%
%
%
%
%
%
%
%
%
%
%
%
%
%
%
%
%
%
%
%
%
%
%
%
%

\begin{table}
\caption{Differential photometry for CPD\,$-57^{\circ}$3509 in the $B$ band.}
\label{tab:photometry_B}
\centering
\begin{tabular}{lcrrr}
\hline
\hline
\multicolumn{1}{c}{MJD}     &
\multicolumn{1}{c}{Airmass} &
\multicolumn{1}{c}{$\Delta m_\mathrm{B}$} &
\multicolumn{1}{c}{$\phi_\mathrm{hyd}$} &
\multicolumn{1}{c}{$\phi_\mathrm{all}$} \\
&
 &
\multicolumn{1}{c}{[mag]} &
 &
 \\
\hline
56754.66893 & 1.216 &    0.0065 & 0.025 & 0.028 \\
56755.71689 & 1.205 & $-$0.0058 & 0.189 & 0.193 \\
56756.62353 & 1.277 & $-$0.0057 & 0.332 & 0.336 \\
56763.67501 & 1.201 & $-$0.0069 & 0.440 & 0.444 \\
56764.59286 & 1.298 &    0.0029 & 0.585 & 0.588 \\
56765.67003 & 1.201 &    0.0020 & 0.754 & 0.758 \\
56766.59765 & 1.274 &    0.0068 & 0.900 & 0.903 \\
56767.65829 & 1.201 & $-$0.0042 & 0.067 & 0.070 \\
56768.59115 & 1.276 &    0.0010 & 0.213 & 0.217 \\
56772.58155 & 1.274 &    0.0071 & 0.841 & 0.844 \\
56773.68069 & 1.213 &    0.0049 & 0.013 & 0.017 \\
56775.66947 & 1.209 & $-$0.0061 & 0.326 & 0.329 \\
56778.58729 & 1.235 &    0.0053 & 0.785 & 0.788 \\
56783.64584 & 1.208 & $-$0.0103 & 0.580 & 0.583 \\
56785.60891 & 1.202 & $-$0.0010 & 0.889 & 0.891 \\
56787.64327 & 1.214 & $-$0.0027 & 0.208 & 0.211 \\
56796.55959 & 1.211 & $-$0.0041 & 0.610 & 0.612 \\
56797.61941 & 1.217 &    0.0043 & 0.777 & 0.779 \\
56805.56942 & 1.201 &    0.0083 & 0.027 & 0.028 \\
56806.50446 & 1.244 & $-$0.0072 & 0.174 & 0.175 \\
57112.53783 & 1.831 & $-$0.0188 & 0.285 & 0.275 \\
57112.59368 & 1.460 & $-$0.0143 & 0.294 & 0.283 \\
57113.53480 & 1.834 & $-$0.0020 & 0.442 & 0.431 \\
57113.55310 & 1.682 & $-$0.0005 & 0.445 & 0.434 \\
57113.58957 & 1.466 & $-$0.0163 & 0.450 & 0.440 \\
57113.60865 & 1.386 & $-$0.0136 & 0.453 & 0.443 \\
57113.62245 & 1.339 & $-$0.0073 & 0.456 & 0.445 \\
57113.64136 & 1.288 & $-$0.0179 & 0.458 & 0.448 \\
57113.65530 & 1.258 & $-$0.0119 & 0.461 & 0.450 \\
57114.53666 & 1.792 & $-$0.0029 & 0.599 & 0.589 \\
57114.59523 & 1.428 &    0.0005 & 0.608 & 0.598\\
57114.68590 & 1.214 & $-$0.0111 & 0.623 & 0.612\\
57115.54074 & 1.735 & $-$0.0004 & 0.757 & 0.747\\
57115.59857 & 1.404 & $-$0.0007 & 0.766 & 0.756\\
57115.68972 & 1.209 &    0.0048 & 0.781 & 0.770\\
\hline
\end{tabular}
\end{table}

\begin{table}
\caption{Differential photometry for CPD\,$-57^{\circ}$3509 in the $V$ band.}
\label{tab:photometry_V}
\centering
\begin{tabular}{lcrrr}
\hline
\hline
\multicolumn{1}{c}{MJD}     &
\multicolumn{1}{c}{Airmass} &
\multicolumn{1}{c}{$\Delta m_\mathrm{V}$} &
\multicolumn{1}{c}{$\phi_\mathrm{hyd}$} &
\multicolumn{1}{c}{$\phi_\mathrm{all}$} \\
 &
 &
\multicolumn{1}{c}{[mag]} &
 &
 \\
\hline
56754.67501 & 1.211 &    0.0004 & 0.025 & 0.029\\
56755.72287 & 1.208 & $-$0.0058 & 0.190 & 0.194\\
56756.62949 & 1.265 & $-$0.0071 & 0.333 & 0.337\\
56763.68112 & 1.201 & $-$0.0062 & 0.441 & 0.445\\
56764.59909 & 1.283 &    0.0040 & 0.586 & 0.589\\
56765.87662 & 1.201 & $-$0.0050 & 0.786 & 0.790\\
56766.60432 & 1.261 &    0.0064 & 0.901 & 0.904\\
56767.66445 & 1.201 &    0.0021 & 0.068 & 0.071\\
56768.59730 & 1.264 & $-$0.0066 & 0.214 & 0.218\\
56772.58781 & 1.261 &    0.0017 & 0.842 & 0.845\\
56775.67570 & 1.213 &    0.0011 & 0.327 & 0.330\\
56778.59372 & 1.226 &    0.0022 & 0.786 & 0.789\\
56783.65240 & 1.212 & $-$0.0061 & 0.581 & 0.584\\
56785.61510 & 1.201 &    0.0036 & 0.890 & 0.892\\
56796.56586 & 1.207 & $-$0.0004 & 0.611 & 0.613\\
56806.51094 & 1.235 & $-$0.0014 & 0.175 & 0.176\\
57112.54384 & 1.777 & $-$0.0144 & 0.286 & 0.276\\
57112.59944 & 1.434 & $-$0.0207 & 0.295 & 0.284\\
57113.54069 & 1.781 & $-$0.0143 & 0.443 & 0.432\\
57113.58178 & 1.504 & $-$0.0218 & 0.449 & 0.439\\
57113.59579 & 1.438 & $-$0.0146 & 0.451 & 0.441\\
57113.61466 & 1.365 & $-$0.0299 & 0.454 & 0.444\\
57113.62857 & 1.321 & $-$0.0217 & 0.456 & 0.446\\
57113.64753 & 1.274 & $-$0.0263 & 0.459 & 0.449\\
57113.66151 & 1.247 & $-$0.0220 & 0.462 & 0.451\\
57114.54293 & 1.740 & $-$0.0176 & 0.600 & 0.590\\
57114.60158 & 1.403 &    0.0112 & 0.609 & 0.599\\
57114.69188 & 1.209 & $-$0.0111 & 0.624 & 0.613\\
57115.60458 & 1.381 & $-$0.0086 & 0.767 & 0.757\\
57115.69598 & 1.205 &    0.0121 & 0.782 & 0.771\\
\hline
\end{tabular}
\end{table}

Tables~\ref{tab:photometry_B} and \ref{tab:photometry_V} present the
photometric data of CPD\,$-57^{\circ}$3509,
where the Modified Julian Date, the airmass of the observation,
the differential photometric values $\Delta m_\mathrm{B}$ and
$\Delta m_\mathrm{V}$, and the phase information $\phi_\mathrm{hyd}$ and
$\phi_\mathrm{all}$ are given.

\label{lastpage}

\end{document}